\documentclass[12pt]{article}
\usepackage{hyperref}
\usepackage{amsmath}
\usepackage{graphicx}
\usepackage{amssymb} 
 \usepackage{xcolor}
\definecolor{myblue}{rgb}{0.14,0.11,0.49}
\definecolor{myred}{rgb}{0.74,0.22,0.15}
\definecolor{mygreen}{rgb}{0.05,0.52,0.42}
\definecolor{myyellow}{rgb}{0.96,0.92,0.13}
\definecolor{myorange}{rgb}{1,0.61,0.36}
\definecolor{mypurple}{rgb}{0.71,0.02,1}

\newcommand{\Mat}[1]{{{\boldsymbol{#1}}}}

\def\be{\begin{equation}}
\def\ee{\end{equation}}
\def\bea{\begin{eqnarray}}
\def\eea{\end{eqnarray}}
\def\bi{\begin{itemize}}
\def\ei{\end{itemize}}
\def\dd{\mathrm{d}}

\date{}

\title{{\bf A solution of the non-uniqueness problem of the Dirac Hamiltonian and energy operators}}

\author{Mayeul Arminjon\\
\small\it Laboratory ``Soils, Solids, Structures, Risks'', 3SR\\ \small\it (CNRS and Universit\'es de Grenoble: UJF, Grenoble-INP),\\\small\it BP 53, F-38041 Grenoble cedex 9, France.}

\begin{document}
\maketitle

\begin{abstract} \noindent 
In a general spacetime, the possible choices for the field of orthonormal tetrads lead (in standard conditions) to equivalent Dirac equations. However, the Hamiltonian operator is got from rewriting the Dirac equation in a form adapted to a particular reference frame, or class of coordinate systems. That rewriting does not commute with changing the tetrad field $(u_\alpha )$. The data of a reference frame F fixes a four-velocity field $v$
, and also fixes a rotation-rate field $\Mat{\Omega }$. It is natural to impose that $u_0=v$. We show that then the spatial triad $(u_p)$ can only be rotating w.r.t. F, and that the title problem is solved if one imposes that the corresponding rotation rate $\Mat{\Xi }$ be equal to $\Mat{\Omega }$---or also, if one imposes that $\Mat{\Xi }=\Mat{0}$. We also analyze other proposals which aimed at solving the title problem.\\


\end{abstract}

\section{Introduction}\label{Introduction}

Dirac's original equation is valid for a Minkowski spacetime in Cartesian coordinates and involves four $4\times 4$ complex {\it constant} matrices, say $\gamma ^{\natural \alpha }$ ($\alpha = 0,...,3$), which play the role of coefficients in this equation. These ``Dirac matrices" generate 
the Clifford algebra associated with the Minkowski metric $\eta^{\alpha\beta}$. I.e., they are assigned to obey the anticommutation relation:
\be \label{Clifford-flat-2}
\gamma ^{\natural \alpha } \gamma ^{\natural \beta  } + \gamma ^{\natural \beta  } \gamma ^{\natural \alpha }   = 2\eta^{\alpha\beta}\,{\bf 1}_4, \quad \alpha ,\beta \in \{0,...,3\}.
\ee
Here the matrix $(\eta^{\alpha\beta}) \equiv \mathrm{diag}(1,-1,-1,-1)$, and ${\bf 1}_4$ is the identity matrix, ${\bf 1}_4 \equiv \mathrm{diag}(1,1,1,1)$. Any two solutions $(\gamma ^{\natural \alpha })$ and $(\gamma ^{\sharp \alpha })$ of the basic relation (\ref{Clifford-flat-2}) should be equally valid sets of coefficients for Dirac's original equation. One can prove that the quantum-mechanical predictions of the Dirac equation in a Minkowski spacetime in Cartesian coordinates are indeed fully independent of the chosen solution $(\gamma ^{\natural \alpha })$ \cite{A40}. \\

In a general spacetime manifold V equipped with a Lorentzian metric $\Mat{g}$, the Dirac equation has to be extended so that the Dirac matrices, say $\gamma ^\mu $ ($\mu =0,...,3$), obey the anticommutation relation with this metric $\Mat{g}$:
\be \label{Clifford}
\gamma ^\mu \gamma ^\nu + \gamma ^\nu \gamma ^\mu = 2g^{\mu \nu}\,{\bf 1}_4, \quad \mu ,\nu \in \{0,...,3\},
\ee
where $(g^{\mu \nu})$ is the inverse matrix of the matrix $(g_{\mu \nu})$ made with the components $g_{\mu \nu}$ of $\Mat{g}$ in a local coordinate system or chart $\chi : X\mapsto (x^\mu )$, defined on an open subset U of V. 
\footnote{\
The Dirac matrices $\gamma ^\mu $ are thus defined locally and depend on the chart. In this paper, we shall not discuss the relation between the Dirac matrices $\gamma ^\mu $, which are the standard object introduced in the physics literature about the Dirac equation (see e.g.  Refs. \cite{BrillWheeler1957+Corr, deOliveiraTiomno1962, ChapmanLeiter1976, HehlNi1990, VarjuRyder1998, Leclerc2006}), and the corresponding global chart-independent object, called ``fundamental spinor-tensor" by Lichnerowicz \cite{Lichnerowicz1964} and ``spinor-tensor $\gamma $" by Dimock \cite{Dimock1982}, which is considered by mathematicians. This relation is discussed in detail in Ref. \cite{A45}, where that object is called ``$\gamma $ field".
} 
Now the metric tensor $\Mat{g}$ depends on the spacetime position $X \in\,$V, thus it is a tensor field. In Eq. (\ref{Clifford}), the components $g^{\mu\nu}$ depend thus on the spacetime position, and so do the Dirac matrices $\gamma ^\mu $. The dependence of the $g^{\mu \nu }$ 's on the spacetime position implies that there is a vast continuum of different possibilities for the $\gamma ^\mu $ field, all being solutions of the basic relation (\ref{Clifford}). In the standard extension of the Dirac equation to a curved spacetime, which is due to Weyl \cite{Weyl1929b} and to Fock \cite{Fock1929b}, and which will be named the ``Dirac-Fock-Weyl" equation ({\it DFW} for short), the $\gamma ^\mu $ field is defined from an orthonormal tetrad field; see Subsect. \ref{tetrad fields} below. The connection used to define the covariant derivatives in the DFW equation depends on the tetrad field in such a way that the equation is covariant under any smooth change of this tetrad field, i.e., under any ``local Lorentz transformation": $X \mapsto L(X) \in {\sf SO(1,3)}$, depending smoothly on $X \in \mathrm{V}$ \cite{BrillWheeler1957+Corr,ChapmanLeiter1976}. (Except for spacetimes that are not simply connected \cite{Isham1978}.)\\

However, the Hamiltonian operator H and the energy operator E, hence the associated energy states, are got from rewriting the Dirac equation in a form adapted to a particular reference frame. (The operator E coincides with the Hermitian part of H \cite{Leclerc2006, A43}.) It is not a priori obvious that this rewriting commutes with changing the coefficients $\gamma ^\mu $ in the Dirac equation. For the Dirac equation in a Minkowski spacetime with constant Dirac matrices $\gamma ^\mu $, in Cartesian (or affine) coordinates, it happens to be the case \cite{A40}, essentially because the change of the matrices $\gamma ^\mu $ does not depend on the spacetime position. For a general spacetime, and in fact already if one applies any ``curved spacetime Dirac equation" to a Minkowski spacetime---even in Cartesian coordinates---the $\gamma ^\mu $ matrices and their admissible changes are allowed to depend on the spacetime position. It follows from this that rewriting the Dirac equation in a form adapted to a particular reference frame does not generally commute with changing the $\gamma ^\mu $ matrices in the Dirac equation. This is the source of an actual non-uniqueness problem for the operators H and E. One of the first manifestations of this problem was observed by Ryder \cite{Ryder2008}. He found that, for spherical coordinates whose azimuth angle $\phi $ expresses a uniform rotation with respect to an inertial frame in a Minkowski spacetime, the presence or absence of Mashhoon's ``spin-rotation coupling term" \cite{Mashhoon1988} in the DFW Hamiltonian depends on the choice of the tetrad field. He ``suggest[ed] that it is a general result that the correct way of defining a rotating frame is with reference to Fermi-Walker transport". He did not mention a non-uniqueness problem, however. Independently of that work, the condition for the general Dirac Hamiltonian operator to be Hermitian was derived in a general curved spacetime in Ref. \cite{A42}. It was found that, for DFW, the validity of this hermiticity condition depends on the choice of the $\gamma ^\mu $ matrices. It was noted that a uniqueness problem thus arises for the DFW equation, even in a flat spacetime \cite{A42}. Thereafter, this problem was studied in detail \cite{A43}. It was thus found that, among the admissible changes which may be imposed to the field of Dirac matrices, the vast majority does not lead to an equivalent Hamiltonian operator H, nor to an equivalent energy operator E, whence a lack of uniqueness; and that even the Dirac energy spectrum is not unique in a given coordinate system in a given spacetime $(\mathrm{V},\Mat{g})$. This was found to apply \cite{A43} to the DFW equation, as also to the two alternative Dirac equations in a curved spacetime, proposed previously \cite{A39}.\\

In two recent papers, Gorbatenko \& Neznamov have proposed an attempt at solving the hermiticity problem and the non-uniqueness problem of the Hamiltonian associated with the standard Dirac equation in a curved spacetime (DFW), first for a stationary gravitational field \cite{GorbatenkoNeznamov2010} and then for a general gravitational field \cite{GorbatenkoNeznamov2011}, by having recourse to pseudo-Hermitian quantum mechanics. They are working in a fixed chart. Let us observe that, for a stationary metric, it is easy to propose a solution of both the hermiticity problem and the non-uniqueness problem: in view of Eq. (\ref{Clifford}), it is then natural to take for $\gamma ^\mu $ field one that does not depend on the time coordinate $t$. This leads to a Hermitian Hamiltonian \cite{Leclerc2006,A42}. Any two time-independent $\gamma ^\mu $ fields are related together by a time-independent local similarity
, hence (\cite{A43}, Eq. (48)) give rise to equivalent Hamiltonian operators. For a general, time-dependent metric, Gorbatenko \& Neznamov \cite{GorbatenkoNeznamov2011} propose to define a unique Hermitian Hamiltonian operator $\mathrm{H}_\eta$. It turns out that $\mathrm{H}_\eta$ is just the energy operator obtained with what they call ``the Schwinger tetrad", i.e., an orthonormal tetrad $(u_\alpha )$, of which each among the three ``spatial" vectors $u_p \ (p=1,2,3)$ has a zero ``time" component, in the chart considered. (In our notation, this writes $a^0_{\ \,p}=0\, (p=1,2,3)$. Such tetrads were considered in Refs. \cite{A43,A42}, and their existence proof was outlined in Footnote 8 of Ref. \cite{A43}. Note that the notion of a Schwinger tetrad depends thus on the chart considered.) Their calculations are correct but, as is shown in Appendix \ref{GorbatNeznam}, their conclusion that ``For any system of tetrad vectors, after the specified operations we will have one and the same Hamiltonian $\mathrm{H}_\eta$" is incorrect. In short, this is because, in a given chart, there are a lot of different tetrads which are a Schwinger tetrad in that chart; and the energy operator in a given chart still depends on the choice of the Schwinger tetrad, just as it depends on the choice of the general tetrad. \\

In the present paper, a viable, physically motivated solution is presented to the non-uniqueness problem of the Dirac Hamiltonian and energy operators. This solution works provided that the $\gamma ^\mu$ field is defined from a tetrad field and that the Dirac equation is covariant under any change of the tetrad field. Both facts apply to the DFW equation. The proposed solution consists in imposing that, at each spacetime point, the local angular velocity tensor of the spatial triad $(u_p)$, as we will define it in this paper, be equal to the local angular velocity tensor of the reference frame, as it has been defined by Weyssenhoff \cite{Weyssenhoff1937} and Cattaneo \cite{Cattaneo1958}. In Section \ref{Non-uniquenessPb}, we summarize the framework and the existing non-uniqueness problem. We emphasize there that both the Hamiltonian operator and the energy operator depend necessarily and naturally on the reference frame, not only for the Dirac equation. In contrast, the non-uniqueness problem
is specific to the curved-spacetime Dirac equation. It occurs in any given reference frame in any spacetime. We also note that the data of a tetrad field is {\it more} than that of a reference frame. In Section \ref{SolvingPb}, we propose our solution. We recall the definition of the local angular velocity tensor of a reference frame \cite{Weyssenhoff1937,Cattaneo1958}. Then we define the local angular velocity tensor of the spatial triad in an orthonormal tetrad. We show that, by restricting the consideration to tetrad fields for which the two rotation rates are equal, one defines unique Hamiltonian and energy operators.

\section{The non-uniqueness problem and its cause}\label{Non-uniquenessPb}

\subsection{The Hamiltonian operator and its dependence on the reference frame}\label{H-reference frame}

Consider a linear differential operator $P$ acting on regular ``wave functions" $\psi $, which are defined over a spacetime manifold V, and whose values $\psi (X)$ at each point $X \in\mathrm{V}$ belong to some vector space $\mathrm{E}_X$ of dimension $m$. (In precise terms: each function $\psi$ is a smooth section of some vector bundle E of dimension $m$, with base V. We assume that $P$ is an internal operator, i.e., $P\psi $ is a smooth section of the same vector bundle E.) By definition, the Hamiltonian operator associated with the linear PDE 
\be\label{P psi=0}
P\psi =0
\ee
in a certain chart: $X\mapsto (x^\mu )$, is a linear differential operator H that: i) contains no derivative with respect to the coordinate time $t\equiv x^0/c$, and ii) is such that (\ref{P psi=0}) rewrites equivalently as the Schr\"odinger equation:
\be \label{Schrodinger-general}
i \frac{\partial \Psi }{\partial t}= \mathrm{H}\Psi, 
\ee
where $\Psi \equiv (\Psi^a)_{a=1,...,m}$ is the column vector made with the components of $\psi $ in some ``frame field", i.e., a smooth field of bases of E: $X \mapsto (e_a(X))$, with $(e_a(X))$ a basis of E$_X$. A such operator H does exist if (\ref{P psi=0}) contains only first-order time derivatives of the components $\Psi ^a$ of $\psi $, in such a way that the corresponding matrix $M$, entering the term $M\Psi_{,0} =(M^a_{\ \,b} \Psi ^b_{,0})$ in the local expression of $P\psi $, is invertible. In that case, the operator H is unique for a given linear equation in a given chart on V, $\chi : X \mapsto {\bf X}\equiv (x^\mu )$, and in a given frame field on E. In a different chart, say $\chi' : X \mapsto {\bf X}' \equiv(x'^\mu )={\bf F}({\bf X})$, this procedure gives rise to a different operator H$'$, which in general is not related simply to H. Usually, the frame field on E is not changed as one changes the chart on V. 
\footnote{\
However, when E is the complexified tangent bundle $\mathrm{T}_{\sf C}\mathrm{V}$, the frame field on it can be taken to be the natural basis $(\partial _\mu )$ associated with the chart. Then, on changing the chart, the frame field is changed to the new natural basis $(\partial' _\mu )$, and $\Psi $ transforms as a true vector \cite{A45}. This corresponds with the ``tensor representation of the Dirac fields" \cite{A42,A39}.
}
Then, the column vector $\Psi$ is left invariant in the change of the chart: $\Psi'({\bf F}({\bf X})) =\Psi ({\bf X})$, hence Eq. (\ref{Schrodinger-general}) transforms as:
\be\label{H'-psi-scalar}
\mathrm{H}'\Psi' \equiv i\frac{\partial \Psi' }{\partial t'} =i\frac{\partial \Psi }{\partial x^\mu }\frac{\partial x^\mu }{\partial t'}=\frac{\partial t }{\partial t' }\mathrm{H}\Psi + i\frac{\partial \Psi }{\partial x^j }\frac{\partial x^j }{\partial t'}.
\ee
It follows that H and H$'$ are independent, inequivalent operators, unless the coordinate change is a purely spatial one \cite{A42}, i.e., unless
\be\label{purely-spatial-change}
x'^0=x^0,\ x'^j=f^j((x^k)).
\ee
In that case, one gets from (\ref{H'-psi-scalar}) the following invariance relation under the change of chart: 
\be\label{H'=H}
\forall {\bf X},\ (\mathrm{H}'\Psi')({\bf F}({\bf X}))=(\mathrm{H}\Psi)({\bf X}). 
\ee
In words: the Hamiltonian acts as a unique operator inside any set of charts which exchange by a purely spatial change (\ref{purely-spatial-change}). This is also true when $\Psi$ transforms as a true vector on a coordinate chart \cite{A42}.\\

The relation (\ref{purely-spatial-change}) between two charts is an equivalence relation for charts which are all defined on a given open set $\mathrm{U} \subset \mathrm{V}$ \cite{A44}. We call {\it reference frame} an equivalence class F of charts for this relation, for a given open set U. Indeed, a such class defines a unique {\it reference fluid} in the sense of Cattaneo \cite{Cattaneo1958}, that is a three-dimensional congruence of time-like world lines (observers): in any chart $\chi $ belonging to a given class (reference frame) F, these world lines are the lines 
\be\label{world-line}
x^j=\mathrm{Constant} \ (j=1,2,3), \quad {\bf X}\equiv (x^\mu) \in \chi (\mathrm{U}).
\ee
It follows from the relation (\ref{purely-spatial-change}) that a world line $x\subset \mathrm{V}$, whose expression is (\ref{world-line}) in a chart belonging to a given class F, has the same expression (with primes, except for U) in any other chart $\chi' \in \mathrm{F}$  \cite{A44}. Let M be the set of the world lines (\ref{world-line}), thus defined from a reference frame F. 
\footnote{\ 
The world-lines (\ref{world-line}) and their set M are left unchanged, not only under a purely spatial change of the coordinates, Eq. (\ref{purely-spatial-change}), but also under a general change of the {\it time} coordinate, $x'^0=f((x^\mu ))$. Thus, our definition of a reference frame as an equivalence class of charts modulo the purely spatial changes (\ref{purely-spatial-change}) involves fixing a reference fluid {\it and} fixing a time coordinate. This additional fixing is imposed by the fact that the Hamiltonian operator (and indeed also a classical Hamiltonian) does depend on the choice of the time coordinate, see Eq. (\ref{H'-psi-scalar}). If we are given only a reference fluid, then the charts in which the world lines of the congruence have the form (\ref{world-line}) are called {\it adapted} to the reference fluid, they exchange by \cite{Cattaneo1958}: 
\be\label{adapted-change}
x'^0=f((x^\mu )),\ x'^j=f^j((x^k)).
\ee
}
Let $ \chi \in \mathrm{F}$ be any chart belonging to the class F. Consider the mapping which associates with any world line $x \in \mathrm{M}$, the triad of its constant spatial coordinates in the chart $\chi $:
 \be\label{def-chi-tilde}
\widetilde{\chi }: \mathrm{M}\rightarrow {\sf R}^3,\quad x\mapsto {\bf x} \equiv (x^j).
\ee
The set M is naturally equipped with a structure of three-dimensional differentiable manifold, for which any mapping $\widetilde{\chi }$, with $\chi \in \mathrm{F}$, is a global chart of M \cite{A44}. We call M the {\it space manifold} associated with the reference frame F.\\

Thus, the Hamiltonian operator naturally depends on the reference frame. The argument given above does not depend on a particular wave equation. This dependence imposes that, if we wish to define a unique Hamiltonian operator for the Dirac equation, then we have first to fix the reference frame. Note that the data of a chart $\chi $, with its domain U (an open set), determines a unique reference frame: the equivalence class F of $\chi $ modulo the purely spatial changes (\ref{purely-spatial-change}). If $\mathrm{U=V}$, the chart $\chi $ and the associated reference frame F are global. There exists a global chart, iff V is diffeomorphic to an open subset of ${\sf R}^4$. This is true even for G\"odel's spacetime, however ``pathological" it may look. However, for the sake of generality, we will not assume that there exists a global chart. The price to pay is that, if $\mathrm{U \ne V}$, a world line (\ref{world-line}) is not necessarily a connected set.

\subsection{Tetrad fields and spin transformations}\label{tetrad fields}

As was indicated in the \hyperref[Introduction]{Introduction}, the non-uniqueness of the Dirac Hamiltonian comes from the fact that there is a vast continuum of different possibilities for the $\gamma ^\mu $ field, all being solutions of the basic relation (\ref{Clifford}). In this paper, we will restrict ourselves to the case where the $\gamma ^\mu $ field is defined from an orthonormal tetrad field $(u_\alpha)$. Given some local chart $\chi : X\mapsto (x^\mu ), \mathrm{U} \rightarrow  {\sf R}^4$, one may express  the vectors $u_\alpha $ in the natural basis $(\partial _\mu )$ as  $u_\alpha =a^\mu_{\ \,\alpha}\,\partial_\mu$. The definition of the $\gamma ^\mu $ field is then written locally (for $X \in\mathrm{U}$) as: 
\be \label{flat-deformed}
 \gamma ^\mu(X) = a^\mu_{\ \,\alpha}(X)  \ \gamma ^{ \natural \alpha},
\ee 
see e.g. Refs. \cite{BrillWheeler1957+Corr,ChapmanLeiter1976}. The orthonormality condition of the tetrad writes
\be\label{orthonormal tetrad}
\Mat{g}(u_\alpha ,u_\beta )= g_{\mu \nu }\,a^\mu _{\ \,\alpha} \,a^\nu _{\ \,\beta } = \eta_{\alpha\beta}.
\ee
Equations (\ref{Clifford-flat-2}) and (\ref{orthonormal tetrad}) imply that the $\gamma ^\mu$ field defined by (\ref{flat-deformed}) satisfies the anticommutation relation (\ref{Clifford}). The orthonormality condition (\ref{orthonormal tetrad}) remains satisfied if we change the tetrad field by applying to the starting one a local (proper) Lorentz transformation
$L: \ \mathrm{V} \rightarrow  {\sf SO(1,3)},\ X \mapsto L(X)$: 
\be\label{ChangeTetrad}
\tilde{u}_\beta  =L^\alpha  _{\ \, \beta  }\,u _\alpha=\tilde{a}^\mu _{\ \, \beta  }\,\partial _\mu,\qquad
\tilde{a}^\mu _{\ \, \beta  }=a^\mu _{\ \, \alpha }\,L^\alpha _{\ \, \beta  }.  
\ee
The new tetrad field $\tilde{u}_\beta $ is associated by Eq. (\ref{flat-deformed}) with a new field of Dirac matrices, $\widetilde{\gamma} ^\mu \equiv \tilde{a}^\mu _{\ \, \beta}  \gamma ^{ \natural \beta }$. Just because that new field also satisfies the anticommutation relation (\ref{Clifford}), we know that it can be deduced from the former one  $\gamma ^\mu $ by a local similarity transformation \cite{Pauli1936}, $X \in \mathrm{V} \mapsto S(X) \in {\sf GL(4,C)}$, which is unique up to a complex scalar $\lambda (X)$ \cite{A40}:
\be \label{similarity-gamma}
\widetilde{\gamma} ^\mu =  S^{-1}\gamma ^\mu S, \quad \mu =0,...,3.
\ee
However, here the local similarity $S$ corresponds to a change of the tetrad field by a local Lorentz transformation, Eq. (\ref{ChangeTetrad}). It follows that, at each spacetime point $X$, we can take the $4 \times 4$ matrix $S(X)$ in the spin group ${\sf Spin(1,3)}$ and such that 
\be\label{L=Lambda o S}
(\forall X \in \mathrm{V}) \quad \Lambda (S(X)) =L(X), 
\ee
where $\Lambda: {\sf Spin(1,3)} \rightarrow {\sf SO(1,3)}$ is the two-to-one covering map of the special Lorentz group by the spin group. See e.g. \cite{Isham1978}. In other words, we can take $S(X)=\pm {\sf S}(L(X))$, where $L\mapsto \pm {\sf S}(L)$ is the spinor representation (which is defined only up to a sign, contrary to the covering map of which it is the ``inverse") \cite{A42}. We will call ``spin transformation" a {\it smooth} local similarity transformation $S$ taking values in ${\sf Spin(1,3)}$. With any spin transformation $S$, Eq. (\ref{L=Lambda o S}) associates a smooth local Lorentz transformation $L$. Conversely, depending on the topology of the spacetime V, all smooth local Lorentz transformations $L$, or  only a subclass of them, can be ``lifted" to a spin transformation $S$ such that $\Lambda \circ S =L$ \cite{Isham1978}.

\subsection{Non-uniqueness of the Dirac Hamiltonian}

The starting Hamiltonian operator H, before application of a local similarity transformation, is got by rewriting the Dirac equation as the Schr\"odinger equation (\ref{Schrodinger-general}). Suppose that some form of the Dirac equation is covariant under the local similarity transformation $S$, which is applied to the Dirac matrices according to Eq. (\ref{similarity-gamma}), and simultaneously to the wave function $\psi $ (or rather to the column matrix $\Psi$ of its components) according to 
\be\label{similarity-psi}
\widetilde{\Psi } = S^{-1}\Psi .
\ee
Then, the new Hamiltonian operator $\widetilde{\mathrm{H}}$ is got by inserting the change of unknown function (\ref{similarity-psi}) into (\ref{Schrodinger-general}) and by rewriting this as a Schr\"odinger equation for $\widetilde{\Psi }$. This gives immediately
\be\label{Htilde}
\widetilde{\mathrm{H}}=S^{-1} \mathrm{H}S-i\,S^{-1}\partial_t S. 
\ee
On the other hand, it is fairly obvious that the two operators H and $\widetilde{\mathrm{H}}$ are mathematically and physically equivalent iff
\be \label{similarity-invariance-H}
\widetilde{\mathrm{H} }  =  S^{-1}\,\mathrm{H}\, S.
\ee
It is indeed easy to prove \cite{A43} that, under this condition only, all scalar products are equal, i.e., for all $\Psi ,\Phi $ in the domain $\mathcal{D}$ of H
\be \label{Hamiltonian-matrix-conserved}
(\widetilde{\mathrm{H} }  \,\widetilde {\Psi} \ \widetilde {\mid }\   \widetilde {\varphi}  )=(\mathrm{H} \, \Psi \mid    \Phi  ) .
\ee
(The relevant scalar product $(\,\mid\,)$, Eq. (\ref{Hermitian-sigma=1-g}) below, is modified by the application of a local similarity transformation, becoming $(\,\widetilde {\mid }\,)$ \cite{A43}.) In particular, the operators H and $\widetilde {\mathrm{H}}$ have then the same spectrum. \\

By comparing (\ref{Htilde}) and (\ref{similarity-invariance-H}), we see that a local similarity transformation $S$, under which the Dirac equation is covariant, provides us with a Hamiltonian operator that is equivalent to the starting one, iff
\be\label{partial_0 S=0}
\partial_t S=0,
\ee
that is, iff the similarity $S$ is independent of the time $t\equiv x^0$---in the reference frame F which is considered, see the end of Subsect. \ref{H-reference frame}. {\it Whence follows the non-uniqueness of the Dirac Hamiltonian} \cite{A43}.\\

In Ref. \cite{A43}, Eqs. (\ref{Htilde}) and (\ref{partial_0 S=0}) were established specifically for the DFW equation, by using the explicit form of the DFW Hamiltonian and its change after a spin similarity. Here, we see that they depend only on the covariance of the Dirac equation under the similarity, or equivalently on the invariance of the corresponding Lagrangian density:
\be\label{Lagrangian-density}
l=\sqrt{-g}\ \frac{i}{2}\left[\,\overline{\Psi}\gamma^{\mu}(D_{\mu}\Psi)-
\left(\overline{D_{\mu}\Psi} \right)\gamma^{\mu}\Psi+2im\overline{\Psi}\Psi\right],
\ee
where $\overline{\Psi}\equiv \Psi ^\dagger A$, with $ \Psi ^\dagger$ denoting the complex conjugate transpose of $\Psi$, and similarly $\overline{D_{\mu}\Psi}\equiv \left(D_{\mu}\Psi \right) ^\dagger A$. (Here, $A$ is the hermitizing matrix \cite{Pauli1936,A40,A42}.) Indeed, the DFW equation as well as all alternative forms of the curved spacetime Dirac equation proposed recently \cite{A39,A45}, all derive from this Lagrangian \cite{A45}. One may check that, in order that the Lagrangian (\ref{Lagrangian-density}) be invariant under the similarity (\ref{similarity-gamma}), (\ref{similarity-psi}), it is necessary and sufficient that the connection matrices $\Gamma _\mu $, which define the covariant derivatives $D_\mu =\partial_\mu +\Gamma _\mu $, transform as
\be\label{Gamma-tilde-psitilde=S^-1 psi}
\widetilde{\Gamma }_\mu = S^{-1}\Gamma _\mu S+ S^{-1}(\partial _\mu S).
\ee
(The matrices $\Gamma _\mu $ for DFW do follow the rule (\ref{Gamma-tilde-psitilde=S^-1 psi}) if the local similarity $S$ is specifically a spin transformation \cite{ChapmanLeiter1976}.) Therefore, if we impose on some form of the Dirac equation that the connection matrices change according to Eq. (\ref{Gamma-tilde-psitilde=S^-1 psi}) after a local similarity, then that form transforms covariantly.
\footnote{\
If the local similarity is a passive one, i.e., if it follows from a change of the frame field on the vector bundle E, then the rule (\ref{Gamma-tilde-psitilde=S^-1 psi}) is satisfied automatically \cite{A45}.
}
So that the new Hamiltonian is (\ref{Htilde}), thus is equivalent to the starting one iff the similarity does not depend on time.

\subsection{Non-uniqueness of the Dirac energy operator}

In the case of time-dependent fields, the Dirac Hamiltonian H is generally not Hermitian \cite{Leclerc2006,A42}. Then, the relevant spectrum is that of the Hermitian part of H:
\be\label{E:=H^s}
\mathrm{E}=\mathrm{H}^{s}\equiv \frac{1}{2}(\mathrm{H}+\mathrm{H}^\ddagger ),
\ee
where the Hermitian conjugate operator $\mathrm{H}^\ddagger$ is w.r.t. the unique scalar product identified in Ref. \cite{A42}:
\be \label{Hermitian-sigma=1-g}
(\Psi  \mid \Phi  ) \equiv \int \Psi^\dagger B \Phi\ \sqrt{-g}\ \dd^ 3{\bf x}, \qquad B\equiv A\gamma ^0.
\ee
The operator E is called the ``energy operator". The justification for this name is that one can show for the general Dirac equation derived from the Lagrangian (\ref{Lagrangian-density}) that the expected value of the operator E equals the classical field energy as derived from the Lagrangian (\ref{Lagrangian-density}), see Ref. \cite{A43}. This has been proved by Leclerc \cite{Leclerc2006} for the DFW equation in the particular case of a ``Schwinger tetrad". From the explicit expression of $\mathrm{E-H}$ \cite{A43}, it is easy to check that the energy operator E transforms like the Hamiltonian H under a purely spatial change (\ref{purely-spatial-change}) of the chart, hence E depends only on the reference frame (for a given $\gamma ^\mu $ field, though accounting for its transformation law on a change of the chart). E.g., when the wave function transforms as a scalar under coordinate changes, E is invariant under a change (\ref{purely-spatial-change}), as in Eq. (\ref{H'=H}) for H.\\

As with the Hamiltonian operator, the energy operator $\widetilde{\mathrm{E}}$ after a local similarity $S$ is physically equivalent to the starting energy operator E, iff
\be \label{similarity-invariance-E}
\widetilde{\mathrm{E} }  =  S^{-1}\,\mathrm{E}\, S.
\ee
The condition on $S$ in order for this to occur, was derived for DFW theory to be \cite{A43}:
\be\label{SEtilde=ES DFW}
B(\partial _0S)S^{-1}-\left[B(\partial _0S)S^{-1}\right]^\dagger \equiv 2 \left[B(\partial _0S)S^{-1}\right]^a = 0.
\ee
This derivation depended only on Eq. (\ref{Htilde}) (Eq. (47) in Ref. \cite{A43}) and on the following one, which results from the explicit expression of the operators H and E for the general Dirac equation \cite{A43}:
\be\label{SEtilde-ES}
\widetilde{\mathrm{E}} -S^{-1}\mathrm{E}S = \widetilde{\mathrm{H}} - S^{-1}\mathrm{H}S+\frac{i}{2}\left [S^{-1}B^{-1}(S^\dagger )^{-1}(\partial _0S)^\dagger BS+ S^{-1}(\partial _0S) \right ].
\ee
Therefore, the condition (\ref{SEtilde=ES DFW}) applies also to the general Dirac equation, provided that it transforms covariantly under the local similarity $S$, i.e., if Eq. (\ref{Gamma-tilde-psitilde=S^-1 psi}) applies. Just like Eq. (\ref{partial_0 S=0}) for H, the condition (\ref{SEtilde=ES DFW}) for the equivalence of E and $\widetilde{\mathrm{E}}$ is generally not satisfied, {\it whence the non-uniqueness of the Dirac energy operator.} Note that any time-independent similarity satisfies (\ref{SEtilde=ES DFW}).

\subsection{Reference frames vs. tetrad fields}

As we have seen in Subsect. \ref{H-reference frame}, the Hamiltonian operator depends necessarily and naturally on the {\it reference frame} F (the meaning of that expression being defined there), not only for the Dirac equation. The non-uniqueness of H and E occurs in any given reference frame F. Thus, when one defines the $\gamma ^\mu $ field from an (orthonormal) tetrad field, the non-uniqueness problem has a relation with the fact that the choice of the tetrad field is not fixed by the choice of F. However, the data of a reference frame, as an equivalence class F of charts modulo the purely spatial changes (\ref{purely-spatial-change}), normally determines a unique four-velocity field $v$. The tangent vector along a world line (\ref{world-line}) is time-like iff $g_{00}>0$ \cite{A44}. Henceforth, {\it we shall consider only charts $\chi$ whose domain $\mathrm{U}$ is such that $g_{00}(X)>0$ for $X \in \mathrm{U}$.} This condition is invariant under a change (\ref{purely-spatial-change}), thus it is imposed on the reference frame F. Under this condition, we define $v$ as the normed tangent vector, whose components in any chart $\chi$ belonging to F are:
\be\label{Vmu}
v^0\equiv \frac{1}{\sqrt{g_{00}}}, \qquad v^j=0.
\ee
These are invariant under the purely spatial changes (\ref{purely-spatial-change}). More generally, the vector with components (\ref{Vmu}) transforms covariantly under the changes (\ref{adapted-change}) of adapted coordinates (when they satisfy $\partial x'^0/\partial x^0 >0$), thus it depends only on the reference fluid associated with F. Thus $v$ is the four-velocity of that reference fluid \cite{Cattaneo1958,A44}. Conversely, the data of an orthonormal tetrad field $(u_\alpha )$ contains that of the four-velocity vector field $u_0$, with $\Mat{g}(u_0,u_0)=1$. Hence it is natural to impose on the tetrad field the condition that the time-like vector of the tetrad is the four-velocity of the reference fluid:
\be\label{u_0=v}
u_0=v.
\ee
This condition was proposed by Mashhoon \& Muench \cite{MashhoonMuench2002} and by Maluf {\it et al.} \cite{MalufFariaUlhoa2007}, although these authors considered $v$ as the 4-velocity of an observer and did not link it to a class of coordinate systems. (Also, what Maluf {\it et al.} \cite{MalufFariaUlhoa2007} call a ``reference frame" is exactly a tetrad field, in contrast with our terminology.)  However, of course, the mere data of $u_0$ does not fix the orthonormal tetrad $(u_\alpha )$. The spatial vectors of the tetrad, $u_p \ (p=1,2,3)$, are generally not ``attached'' to F. We will show that the spatial triad $(u_p)$ as a whole can only be {\it rotating} w.r.t. F, and we shall fix this rotation by fixing the corresponding angular velocity tensor.

\section{A solution of the non-uniqueness problem}\label{SolvingPb}

\subsection{Angular velocity tensor field of a reference frame}\label{Omega}

Weyssenhoff \cite{Weyssenhoff1937} introduced an antisymmetric spatial tensor field:
\be\label{Weyssenhoff}
\omega _{jk}\equiv \partial _jg_k-\partial _kg_j-(g_j\partial _0g_k-g_k\partial _0g_j) \quad (j,k=1,2,3), \qquad g_j\equiv \frac{g_{0j}}{g_{00}}.
\ee
One checks easily that $\omega _{jk}$ is indeed a spatial $(0 \ 2)$ tensor field, i.e., it transforms as does a usual $(0 \ 2)$ tensor, for purely spatial changes of the chart, Eq. (\ref{purely-spatial-change}). 
\footnote{\ 
Given a reference frame F, covering some open subset U of the spacetime (Subsect. \ref{H-reference frame}), we can define a spatial tensor field (depending, however, on the {\it spacetime} position $X \in \mathrm{U}$) as a tensor field on the space manifold M associated with F. E.g., a spatial vector field can be defined as a mapping ${\bf u}$ from $\mathrm{U}$ into the tangent bundle $\mathrm{TM}$, such that $\forall X \in \mathrm{U},\ {\bf u}(X) \in \mathrm{TM}_x$, the tangent space to M at $x$ --- where, $\forall X \in \,$U, $x=x(X)$ is the unique world line of M, such that $X \in x$  \cite{A44}. One defines in the same way a spatial tensor field of any type $(p \ q)$, by substituting for TM the appropriate tensor product $\mathrm{TM} \otimes ...\otimes \mathrm{TM} \otimes \mathrm{TM}^\circ \otimes ... \otimes \mathrm{TM}^\circ $ with $\mathrm{TM}^\circ$ the cotangent bundle. All spatial tensor fields discussed in this paper, including the tensor (\ref{Weyssenhoff}), are also ones in this geometrical sense.
}
Weyssenhoff introduced it in order to characterize a ``normal congruence", i.e., a reference fluid for which there exists adapted coordinates (Footnote 3) such that $g_{0j}=0 \ (j=1,2,3)$: this is the case iff the Weyssenhoff tensor (\ref{Weyssenhoff}) vanishes \cite{Weyssenhoff1937, Cattaneo1958}. Thus, in that case, we can get $g_{0j}=0$ by a change (\ref{adapted-change}), in fact by a mere change of the time coordinate (synchronization). However, Cattaneo \cite{Cattaneo1958} showed that the tensor (\ref{Weyssenhoff}) has also a precise {\it dynamical} meaning. He rewrote the geodesic equation of motion, in the most general case, in the form of Newton's second law with the proper time $\tau $ of the local observer, the velocity $v^k$, the (velocity-dependent) mass $m$, etc., as measured in a given reference frame with local clocks and rods. In the dynamical equation thus obtained (\cite{Cattaneo1958}, Eq. (101)), the tensor (\ref{Weyssenhoff}) contributes a term which has precisely the form of the Coriolis force of Newtonian theory: $F_{j\, \mathrm{Coriolis}}=-2m\Omega_{jk}v^k$, where the spatial tensor
\be\label{Weyssenhoff modified}
\Omega_{jk} \equiv \frac{1}{2}c\sqrt{g_{00}}\ \omega_{jk} \equiv \frac{1}{2}c\sqrt{g_{00}}\,(\partial _j g_k-\partial _k g_j-g_j\partial _0 g_k+g_k\partial _0 g_j)
\ee
plays thus exactly the role played by the angular velocity tensor of a rotating frame in Newtonian theory. This gives a strong argument for calling $\Omega_{jk}$ the {\it angular velocity tensor field of the reference frame} F. Note that the spatial tensor field $\Omega_{jk}$ depends indeed on the reference frame, i.e., it depends on the reference fluid and on the time coordinate. \\

As an illustrative example, let us calculate this tensor for the metric which is obtained in a Minkowski space for a uniformly rotating reference frame. That is, starting from an inertial reference frame F$'$ defined by Cartesian coordinates $t',x',y',z'$, in which the Minkowski metric has the standard form
\be\label{Minkowski in Cartesian}
ds^2=c^2\,dt'^2-(dx'^2+dy'^2+dz'^2),
\ee
we go from F$'$ to a reference frame F having a uniform rotation with constant angular velocity $+\omega $ around the $z'$ axis w.r.t. F$'$ (we mean constant in terms of the coordinate time of F, $t=t'$), by setting
\be\label{rotating Cartesian-1}
t=t',\ x=x'\cos \omega t + y' \sin \omega t,\ y=-x' \sin \omega t + y' \cos \omega t,\ z=z'.
\ee
The Minkowski metric writes in the coordinates $t,x,y,z$ [see Eq. (\ref{Minkowski in rotating Cartesian-2})]:
\be\label{Minkowski in rotating Cartesian}
ds^2=\left[1-\left(\frac{\omega \rho }{c}\right)^2\right](dx^0)^2+2\frac{\omega }{c} (y\,dx-x\,dy)dx^0-(dx^2+dy^2+dz^2),
\ee
where $\rho  \equiv \sqrt{x^2+y^2}$ and $x^0=ct$. The coordinates $t,x,y,z$, restricted to the domain $\mathrm{U}\equiv \{X \in \mathrm{V}; \omega \rho <c\}$, define the class F. We have thus here:
\be
(g_j)=\frac{\omega }{c} \left( \frac{y}{1-\frac{\omega ^2 \rho ^2}{c^2}}\,,\ \frac{-x}{1-\frac{\omega ^2 \rho ^2}{c^2}}\,,\ 0\right), 
\ee
whence
\be\label{Omega uniformly rotating frame}
\omega _x \equiv \Omega_{32}=0,\quad  \omega _y \equiv \Omega_{13}=0,\quad \omega _z \equiv \Omega_{21} = +\omega /\left[1-\frac{\omega ^2 \rho ^2}{c^2}\right]^\frac{3}{2}.
\ee
Thus we find that the angular velocity vector field of the uniformly rotating reference frame is the constant ``nominal" angular velocity vector $\Mat{\omega} \equiv \omega \partial_z $, up to a second-order term. The presence of that term is due to the fact that the locally measured time and distances are influenced by special-relativistic effects related with the non-uniform velocity generated by the rotation.

\subsection{Angular velocity tensor field of a spatial triad}

Consider a smooth curve in spacetime, $C:\,\xi \in \mathrm{I}  \mapsto C(\xi )=X \in \mathrm{V}$, with tangent vector $T(\xi )\equiv \frac{dC}{d\xi }$,  where I is an open subset of the real line. (The set I, hence also the set $C(\mathrm{I})\subset \mathrm{V}$, is not necessarily connected.) Let $U$ be a smooth vector field defined in a neighborhood of the range $C(\mathrm{I})$. The absolute derivative of $U$ along $C$ is the vector $(DU/d\xi )_C  \equiv D_T\, U \equiv (DU)(T)$, where $D$ is the Levi-Civita connection associated with the metric $\Mat{g}$.
\footnote{\
A connection $D$ on TV associates with any vector field $U$ (i.e., with any local section of TV), a local section $DU$ of the tensor product bundle $\mathrm{TV^\circ \otimes TV\cong Hom(TV,TV)}$ \cite{ChernChenLam1999}. Thus, with any vector field $T$, $DU$ associates a vector field $(DU)(T)$. The latter is the contracted product of $DU$ and $T$ corresponding with the unique dual pair $(\mathrm{TV^\circ,TV})$.
}
As is well known, given any tetrad field $(u_\alpha )$, any connection $D$ on TV is characterized by coefficients $\gamma ^\epsilon _{\ \, \alpha \beta }$, such that
\be\label{connection gamma}
D_\beta \,u_\alpha \equiv (Du_\alpha)(u_\beta )=\gamma ^\epsilon _{\ \, \alpha \beta }\, u_\epsilon .
\ee
Decomposing the vector field $U$ on the tetrad: $U=U^\alpha u_\alpha $, we get
\be\label{absolute derivative}
\left(\frac{DU}{d\xi }\right )_C =dU^\alpha (T)u_\alpha+U^\alpha \times (Du_\alpha)(T )  = \left[ \frac{dU^\epsilon}{d \xi  }+\gamma ^\epsilon _{\ \, \alpha \beta }\,U^\alpha  T^\beta  \right] \,u_\epsilon ,
\ee
just as in the particular case of a coordinate (holonomic) basis, in which case the coefficients $\gamma ^\epsilon _{\ \, \alpha \beta }$ in Eq. (\ref {connection gamma}) are the Christoffel symbols. (We have $\frac{dU^\epsilon }{d \xi  } \equiv dU^\epsilon (T)=\frac{\partial U^\epsilon}{\partial x^\mu }\frac{dx^\mu }{d \xi  } $.) As is well known, $(DU/d\xi)_C$ depends only on $C$ and on the mapping $\xi \mapsto U(\xi )\equiv U(C(\xi ))$ defined on I. Note that the parameter $\xi $ used to define the curve $C$ is arbitrary. Since the Levi-Civita connection is metric-compatible: $D\Mat{g}=0$, the absolute derivative obeys the Leibniz rule:
\be\label{Leibniz for absolute}
\left(\frac{d}{d\xi }\right)_C\,\left(\Mat{g}(U,U')\right)=\Mat{g}\left(\left(\frac{DU}{d\xi }\right )_C,U'\right)+\Mat{g}\left(U,\left(\frac{DU'}{d\xi }\right )_C\right).
\ee
For a tetrad field $(u_\alpha )$, let us decompose the vectors $Du_\alpha /d\xi $ on the tetrad itself:
\be\label{Phi tensor}
\left(\frac{Du_\beta }{d\xi }\right)_C  = \Phi ^\alpha _{\ \,\beta }\, u_\alpha .
\ee
Thus, we have for an orthonormal tetrad $(u_\alpha )$:
\be\label{Phi ST}
\Mat{g}\left(u_\alpha ,\left(\frac{Du_\beta }{d\xi }\right)_C \right)=\Phi ^\epsilon  _{\ \,\beta   }\,\eta _{\alpha \epsilon } \equiv \Phi _{\alpha \beta }.
\ee
From Eq. (\ref{Leibniz for absolute}) with $U=u_\alpha $ and $U'=u_\beta $, we see that this is an antisymmetric tensor: $\Phi _{\beta \alpha } + \Phi _{\alpha \beta }=0 $. At this stage, this tensor is defined over the curve $C$, i.e., for $X=C(\xi)$ with $\xi  \in \mathrm{I}$. \\

We consider henceforth the case that the curve $C$ is a trajectory of the vector field $v \equiv u_0$, the time-like vector of a global tetrad field $(u_\alpha )$ \cite{MashhoonMuench2002,Mashhoon2003,MalufFariaUlhoa2007}. Hence $v$ is a global 4-velocity vector field, thus its trajectories make a 3-D congruence of world lines, or reference fluid. We assume that there is at least one chart $\chi: X \mapsto (x^\mu)$, with domain U (with $g_{00}>0$ in U), which is adapted to that reference fluid. That is, the world lines of the congruence are described in the chart $\chi $ by Eq. (\ref{world-line}). The set of these world lines makes then a space manifold M, see Subsect. \ref{H-reference frame}. (The equivalence class of the chart $\chi$ modulo the relation (\ref{purely-spatial-change}) is a reference frame F.) Then $\Phi _{\alpha \beta }$ becomes a tensor field defined on U: given $X \in \mathrm{U}$, let $x(X)$ be the unique world line $x \in \mathrm{M}$, such that $X \in x$  \cite{A44}. I.e., $x(X)$ is the trajectory of $v$ which passes at $X$, more exactly it is the intersection of that trajectory with the domain U. The tensor $\Phi _{\alpha \beta }(X)$ is defined by taking as curve $C$ in Eq. (\ref{Phi ST}), the world line $x(X)$, parameterized by the time coordinate:
\be\label{time coordinate}
t(Y) \equiv x^0(Y)/c,
\ee
say $C: t\mapsto C(t)=Y\in x(X)$. Hence the tangent vector is
\be\label{w=v ds/dt}
T(t) = \frac{dC}{dt} = \frac{ds}{dt}\,v(C(t)).
\ee
From (\ref{connection gamma}), (\ref{Phi ST}) and (\ref{w=v ds/dt}), we get directly
\be
\Phi _{\alpha \beta }\equiv \Mat{g}\left(u_\alpha ,\left(\frac{Du_\beta }{dt }\right)_C \right) = c \frac{d\tau }{dt} \Mat{g}\left(u_\alpha ,(Du_\beta )(u_0) \right) = c \frac{d\tau }{dt}\eta _{\alpha \epsilon }\,\gamma ^\epsilon _{\ \, \beta 0}, 
\ee
whence the simple expression (confirming the antisymmetry $\Phi _{\beta \alpha } =- \Phi _{\alpha \beta }$, since for an orthonormal tetrad and a metric-compatible connection, the coefficients (\ref{connection gamma}) verify $\gamma _{\alpha \beta \epsilon }=-\gamma _{\beta \alpha \epsilon }$):
\be\label{Phi explicit}
\Phi _{\alpha \beta }= c \frac{d\tau }{dt}\gamma _{\alpha \beta 0}.
\ee
In particular, the field $X \mapsto \Phi _{\alpha \beta }(X)$ is smooth. It depends of course on the orthonormal tetrad field $(u_\alpha )$, and also on the choice of the time coordinate $X \mapsto t(X) \equiv x^0(X)/c$, or rather on the rate $d\tau /dt$, where $d\tau =ds/c$ is the proper time increment along the world line $x(X) \in \mathrm{M}$---that is, the proper time of the local observer bound with the reference fluid having velocity $v$. \\

It may seem natural to use only that proper time to define the tensor $\Phi _{\alpha \beta }$, as is done in Ref. \cite{MalufFariaUlhoa2007}. And indeed, whether or not there exists a chart $\chi $, in which the trajectories of the 4-velocity field $v\equiv u_0$ are described by Eq. (\ref{world-line}), one may use the definition (\ref{Phi ST}) with $C$ being that trajectory of $v$ which passes at $X$, parameterized by the proper time $\tau $. This gives our formula (\ref{Phi explicit}) with $d\tau /dt=1$. However, a priori, the trajectory of $v$ passing at $X$ is defined only in some interval $\mathrm{I}_X$, enclosing the initial time $\tau _0$ such that $C(\tau _0)=X$, and there is no differential structure on the set of these trajectories. \hyperref[Theorem1]{Theorem 1} in Appendix A needs that there exists a chart $(\chi,\mathrm{U}) $, in which the trajectories of  $v$ are described by Eq. (\ref{world-line}). (In the application right below to the non-uniqueness problem of the Dirac Hamiltonian and energy operators, this existence is guaranteed because we start from a given reference frame.) Under this condition, the set M of the world lines (\ref{world-line}) is naturally equipped with a structure of differential manifold \cite{A44}. This allows us to define an (antisymmetric) {\it spatial tensor field} $\Mat{\Phi}$, i.e., a mapping $X \in \mathrm{U} \mapsto \Mat{\Phi}(X) \in \mathrm{TM}^\circ_x \otimes \mathrm{TM}^\circ_x$ with $x=x(X)$, whose components are the $\Phi _{pq}$ components $(p,q=1,2,3)$ of the tensor field $\Phi _{\alpha \beta }$. Otherwise, this subset of components would not have a definite meaning. We show that $\Mat{\Phi}$ is the {\it opposite} of the angular velocity tensor of the spatial triad $({\bf u}_p)$ associated with the tetrad $(u_\alpha )$, Eq. (\ref{Xi _pq}) with $\xi =t$.

\subsection{The proposed solution: equating the two rotation rates}

The data of a reference frame F determines a unique four-velocity field $v$, Eq. (\ref{Vmu}), and also fixes the modified Weyssenhoff angular velocity tensor $\Mat{\Omega }$, Eq. (\ref{Weyssenhoff modified}), which is a spatial tensor. In addition to the condition $u_0=v$, Eq. (\ref{u_0=v}), it is natural to impose on the tetrad field $(u_\alpha )$ the condition that the angular velocity tensor $\Mat{\Xi}$ of the spatial triad $({\bf u}_p)$, Eq. (\ref{Xi _pq}), be equal to $\Mat{\Omega }$:
\be\label{Xi=Omega}
\Mat{\Xi}=-\Mat{\Phi}=\Mat{\Omega }.
\ee
(According to the definition in Appendix \ref{Proof1}: for a given index $p \in \{1,2,3\}$, ${\bf u}_p$ is the spatial vector whose components, in any chart $\chi \in \mathrm{F}$, are simply the spatial components $a^j_{\ \, p} \ (j=1,2,3)$ of the 4-vector $u_p$ in the chart $\chi $.) We now show that this provides a solution to the non-uniqueness problem.

\paragraph{Theorem 2.}\label{Theorem2} {\it In the spacetime $(\mathrm{V},\Mat{g})$, let $(u_\alpha )$ and $(\widetilde{u}_\alpha )$ be two orthonormal direct tetrad fields such that $u_0=\widetilde{u}_0$. Then:} (i) {\it The two tetrad fields exchange by a local Lorentz transformation: $\widetilde{u}_\beta =L^\alpha _{\ \, \beta }\,u_\alpha $, whose matrix has the form} 
\be\label{L=1 tensor R}
L=\begin{pmatrix} 
1 & 0  & 0 & 0\\
0  &  &  & \\
0 &  & R &\\
0 &   &  &
\end{pmatrix}\
\ee
{\it where $R$ is a rotation matrix, $R=R(X) \in {\sf SO(3)}$.} (ii) {\it Assume that, in some reference frame $\mathrm{F}$ with 4-velocity $v=u_0=\widetilde{u}_0$, the corresponding spatial triad fields $({\bf u}_p)$ and $(\widetilde{{\bf u}}_r)$ 
 have the same angular velocity tensor:}
\be\label{Phi=Phi tilde}
\Mat{\Xi} \equiv \Xi _{p q}\,\theta^p \otimes \theta^q = \widetilde{\Xi} _{r s} \,\widetilde{\theta}^r \otimes \widetilde{\theta}^s,
\ee
{\it where $(\theta^p)$ and $(\widetilde{\theta}^r)$ are the bases of spatial one-forms dual to $({\bf u}_p)$ and $(\widetilde{{\bf u}}_r)$, respectively. Then, the matrix $R$ does not depend on the time coordinate: $\partial _t R=0$, thus $R=R(x),\ x \in \mathrm{M}$.} (iii) {\it Therefore, the fields of Dirac matrices $\gamma ^\mu $ and $\widetilde{\gamma }^\mu $ deduced respectively from $(u_\alpha )$ and $(\widetilde{u}_\alpha )$ [Eq. (\ref{flat-deformed})] exchange by a time-independent local similarity $S=S(x)$, hence give rise, in the reference frame $\mathrm{F}$, to equivalent Dirac Hamiltonian operators $\mathrm{H}$ and $\mathrm{\widetilde{H}}$, as well as to equivalent Dirac energy operators $\mathrm{E}$ and $\mathrm{\widetilde{E}}$.}\\

\noindent {\it Proof.} By imposing that $u_0=\widetilde{u}_0$, we get of course that the (local) Lorentz transformation $L$ with $\widetilde{u}_\beta =L^\alpha _{\ \, \beta }\,u_\alpha $ satisfies $L^\alpha _{\ \, 0 }=\delta ^\alpha _0 $, similarly we have $\left(L^{-1}\right)^\alpha _{\ \, 0 }=\delta ^\alpha _0 $. Since $L$ is a Lorentz transformation, we have $L^T \eta L=\eta $, or $L^T=\eta L^{-1}\eta^{-1} $, whence
\be
L^0 _{\ \, \alpha  }=(L^T)_\alpha ^{\ \,0}=\eta _{\alpha \epsilon }\left(L^{-1}\right)^\epsilon  _{\ \, \zeta  }\,\eta ^{\zeta 0}=\eta _{\alpha \epsilon }\left(L^{-1}\right)^\epsilon  _{\ \, 0  }= \eta _{\alpha \epsilon }\delta ^\epsilon _0=\eta _{\alpha 0 }=\delta ^0_\alpha ,
\ee
hence the matrix $L$ has the form (\ref{L=1 tensor R}). It follows that $R^T=R^{-1}$, thus $R \in {\sf O(3)}$. We have $\mathrm{det}\,L=1$, because both $(u_\alpha )$ and $(\widetilde{u}_\alpha )$ are direct. It then follows from (\ref{L=1 tensor R}) that $\mathrm{det}\,R=1$, hence $R \in {\sf SO(3)}$. This proves (i).\\

The time evolution of each triad is governed by Eq. (\ref{Xi tensor}) with $\xi =t$, the time coordinate common to any chart $\chi \in \mathrm{F}$; thus
\be\label{u_p dot, u tilde_r dot}
\frac{\delta {\bf u}_q }{dt }  = \Xi ^p _{\ \,q }\, {\bf u}_p,\qquad \frac{\delta {\bf \widetilde{u}}_s}{dt }  = \widetilde{\Xi} ^r_{\ \, s}\, {\bf \widetilde{u}}_r.
\ee
From (\ref{L=1 tensor R}) and the definition (\ref{Def bf u_p}) of the spatial triad, we have
\be\label{u tilde = R u}
{\bf \widetilde{u}}_s(X)\equiv i_X(\widetilde{u}_s(X)) = i_X(R^q _{\ \,s}(X)u_q(X))=R^q _{\ \,s}(X){\bf u}_q(X),
\ee
whence by (\ref{u_p dot, u tilde_r dot}):
\be\label{u tilde_r dot=}
\widetilde{\Xi} ^r_{\ \, s}\, {\bf \widetilde{u}}_r = (\partial _t R)^q _{\ \,s}\,{\bf u}_q+R^q _{\ \,s}\,\Xi ^p _{\ \,q }\, {\bf u}_p \quad (s=1,2,3).
\ee
We did not use the assumption (\ref{Phi=Phi tilde}) until now. In view of (\ref{u tilde = R u}), that equality of tensors means that the components $\Xi _{pq}=\Xi ^p_{\ \, q}$ and $\widetilde{\Xi} _{rs}=\widetilde{\Xi} ^r_{\ \, s}$ exchange by
\be
\widetilde{\Xi} ^r_{\ \, s}=\left(R^{-1}\right)^ r_{\ \,p}\,\Xi ^p_{\ \, q}\,R^q _{\ \,s}.
\ee 
From this and (\ref{u tilde_r dot=}), we get:
\be
\left(R^{-1}\right)^ r_{\ \,p}\,\Xi ^p_{\ \, q}\,R^q _{\ \,s}{\bf \widetilde{u}}_r = \Xi ^p_{\ \, q}\,R^q _{\ \,s}\,{\bf u}_p = (\partial _t R)^q _{\ \,s}\,{\bf u}_q+R^q _{\ \,s}\,\Xi ^p _{\ \,q }\, {\bf u}_p,
\ee 
whence 
\be
(\partial _t R)^q _{\ \,s}\,{\bf u}_q = {\bf 0} \quad (s=1,2,3),
\ee
hence $\partial _t R=0$, in any chart $\chi \in \mathrm{F}$. That is, the expression in the chart $\chi \in \mathrm{F}$, $\breve{R}_\chi : {\bf X} \mapsto R(\chi^{-1} ({\bf X}))$, of the matrix function on V: $X \mapsto R(X)$, depends only on the spatial coordinates---for any chart $\chi \in \mathrm{F}$. This means, as is easy to check, that $R$ is a function only of $x \in \mathrm{M}$, $R(X)=\hat{R}(x(X))$. This proves (ii). \\

The fields of Dirac matrices $\gamma ^\mu $ and $\widetilde{\gamma }^\mu $ deduced respectively from $(u_\alpha )$ and $(\widetilde{u}_\alpha )$ [Eq. (\ref{flat-deformed})] exchange by a time-independent local similarity $S(x)=\pm {\sf S}(L(x))$ where $L$ has the form (\ref{L=1 tensor R}) with $R=R(x)\ (x\in \mathrm{M})$. It then follows from condition (\ref{partial_0 S=0}) [or (\ref{SEtilde=ES DFW})], that the corresponding Hamiltonian operators $\mathrm{H}$ and $\mathrm{\widetilde{H}}$ [and the corresponding energy operators $\mathrm{E}$ and $\mathrm{\widetilde{E}}$] are equivalent. (Note that we do not need that $S$ depend smoothly on the spatial coordinates, hence no topological restriction on the spacetime needs to be imposed.) This proves point (iii), thus completing the proof of \hyperref[Theorem2]{Theorem 2}.

\section{Discussion and examples}\label{Discussion}

Actually, \hyperref[Theorem2]{Theorem 2} provides us with {\it two} natural solutions to the non-uniqueness problem exposed in Section \ref{Non-uniquenessPb}: ({\bf i}) we may impose that the spatial triad has the same angular velocity tensor as the given reference frame, Eq. (\ref{Xi=Omega}). Or ({\bf ii}) alternatively we may impose simply that the spatial triad has {\it zero} angular velocity tensor, $\Mat{\Xi}=\Mat{0}$. Note that, to use Theorem 2, we have also to impose the condition $u_0=v$, hence either solution is valid only in reference frames associated to one fixed reference fluid, given by its four-velocity field $v$. (Such reference frames may differ by the time coordinate, see Note 3.) That is, if two tetrad fields $(u_\alpha )$ and $(\widetilde{u}_\alpha )$ give rise to the same angular velocity tensor field $\Mat{\Xi}=\Mat{\widetilde{\Xi}}$ in a reference frame F having four-velocity field $v=u_0=\widetilde{u}_0$, then in a reference frame F$'$ having a different four-velocity field $v'$ they have no reason to lead to equivalent Dirac Hamiltonians nor to equivalent Dirac energy operators. 

\paragraph{To compare}\label{Comparison}the solutions ({\bf i}) and ({\bf ii}), let us fix the reference frame and consider two tetrad fields $(u_\alpha )$ and $(\widetilde{u}_\alpha )$: the first one, got from the prescription ({\bf i}), has rotation rate $\Mat{\Xi}=-\Mat{\Omega }$. The second one, following prescription ({\bf ii}), has rotation rate $\Mat{\widetilde{\Xi}}=\Mat{0}$. [Both could be obtained concretely by integrating Eq. (\ref{u_p dot, u tilde_r dot}).] From (\ref{u tilde_r dot=}), we get
\be
\Mat{0} = (\partial _t R)^q _{\ \,s}\,{\bf u}_q-R^q _{\ \,s}\,\Omega ^p _{\ \,q }\, {\bf u}_p \quad (s=1,2,3),
\ee
hence
\be
\partial _t R = \Omega R, \qquad \Omega \equiv (\Omega^p _{\ \,q}).
\ee
Obviously this leads, by integration on a given world line $x \in \mathrm{M}$, to a time-dependent rotation $R=R(t,x)$. For instance, if the rotation rate field of the reference frame satisfies $\partial _t\Omega=0$, we get simply
\be\label{Omega constant}
R(t,x)=\exp\left[(t-t_0)\Omega(x)\right] R_0(x) \qquad (R(t_0,x)=R_0(x)).
\ee
From point (i) in \hyperref[Theorem2]{Theorem 2}, we know that the two tetrad fields $(u_\alpha )$ and $(\widetilde{u}_\alpha )$ exchange by a local Lorentz transformation of the special form (\ref{L=1 tensor R}). Hence the fields of Dirac matrices $\gamma ^\mu $ and $\widetilde{\gamma }^\mu $ deduced respectively from $(u_\alpha )$ and $(\widetilde{u}_\alpha )$ [Eq. (\ref{flat-deformed})] exchange by a {\it time-dependent} local similarity $S(t,x)=\pm {\sf S}(L(t,x))$, with $L$ deduced from $R$ by Eq. (\ref{L=1 tensor R}). Therefore, the corresponding Hamiltonian operators $\mathrm{H}$ and $\mathrm{\widetilde{H}}$, at least, are definitely {\it not} equivalent --- and the corresponding energy operators $\mathrm{E}$ and $\mathrm{\widetilde{E}}$ have no reason to satisfy the condition of equivalence (\ref{SEtilde=ES DFW}). Thus, the two prescriptions ({\bf i}) and ({\bf ii}) are definitely non-equivalent.\\

This applies to the case of a uniformly rotating reference frame F in a Minkowski spacetime: F is the equivalence class of the coordinate chart $\chi : X \mapsto (x^\mu )$ --- the latter being deduced from a global inertial chart $\chi ': X \mapsto (x'^\mu )$ by the coordinate change (\ref{rotating Cartesian-1}), restricted to the domain $\mathrm{U}\equiv \{X \in \mathrm{V}; \omega \sqrt{x'^2+y'^2} <c\}$. The angular velocity tensor $\Mat{\Omega}$ of that reference frame is given by Eq. (\ref{Omega uniformly rotating frame}). It is thus constant at fixed $x^1,x^2,x^3$, which means it is constant on any world line $x$ that belongs to the corresponding space manifold $\mathrm{M}$. (This should not be confused with the first spatial coordinate $x=x^1$.) Hence the foregoing paragraph and its Eq. (\ref{Omega constant}) apply, showing that the two prescriptions ({\bf i}) and ({\bf ii}) are not equivalent. It is likely that the prescription ({\bf i}) lead to the presence of a spin-rotation coupling term \cite{Mashhoon1988} in the energy operator, and that the prescription ({\bf ii}) lead to predicting no such term.\\

Similarly, consider the two tetrad fields studied by Ryder \cite{Ryder2008}. The first one is given by his Eq. (3):
\be\label{Ryder 3}
u_0=\frac{1}{c}\frac{\partial }{\partial t}+\frac{\omega y}{c}\frac{\partial }{\partial x}-\frac{\omega x}{c}\frac{\partial }{\partial y}, \quad u_1=\frac{\partial }{\partial x}, \quad u_2=\frac{\partial }{\partial y}, \quad u_3=\frac{\partial }{\partial z},
\ee
and this equation is written precisely in the ``rotating Cartesian coordinates" $(x^\mu )=(ct,x,y,z)$ got by the change (\ref{rotating Cartesian-1}) here (see Appendix \ref{RotFlat}). Note that, by definition, the four-velocity $v$ of the reference frame F defined by the coordinates $(x^\mu )=(ct,x,y,z)$, is given in those coordinates by Eq. (\ref{Vmu}). From (\ref{Ryder 3})$_1$, it is clear that $v\ne u_0$. As noted by Ryder, the only non-zero coefficients of the Levi-Civita connection, as expressed in that tetrad field $(u_\alpha )$, are:
\be\label{Ryder 7}
\gamma _{120}=-\gamma _{210}=\frac{\omega }{c}.
\ee
Therefore, the angular velocity tensor field $\Mat{\Xi}=-\Mat{\Phi } $ of the spatial triad $({\bf u}_p)$ taken from the tetrad field $(u_\alpha )$ is [Eq. (\ref{Phi explicit})]:
\be\label{Xi Ryder1}
\Xi_{32}=0,\quad  \Xi_{13}=0,\quad \Xi_{21} = +\omega \frac{d\tau }{dt},
\ee
in any reference frame F$''$, whose 4-velocity is $v=u_0$, thus necessarily F$''\,\ne\,$F. In fact, Eq. (\ref{Minkowski tetrad in rotating Cartesian-0}) below shows that we can take for F$''$, the {\it inertial frame} F$'$. Ryder's \cite{Ryder2008} second tetrad field is [his Eq. (10)]:
\be\label{Ryder 10}
\widetilde{u}_0=\frac{1}{c}\frac{\partial }{\partial t}-\frac{\omega }{c}\frac{\partial }{\partial \phi }, \quad \widetilde{u}_1=\frac{\partial }{\partial r}, \quad \widetilde{u}_2=\frac{1}{r}\frac{\partial }{\partial \theta }, \quad \widetilde{u}_3=\frac{1}{r\sin\theta }\frac{\partial }{\partial \phi }.
\ee
It is expressed in ``rotating spherical coordinates" $t,r,\theta ,\phi $ deduced from spherical coordinates $t',r',\theta' ,\phi' $ bound to the inertial frame F$'$, by $t=t',r=r',\theta =\theta',\phi =\phi '-\omega t$, leading to the expression of the Minkowski metric in the coordinates $t,r,\theta ,\phi $, Eq. (8) of Ref. \cite{Ryder2008}. Hence we have
\bea\label{rotating spherical}
x & = & r\sin\theta \cos\phi =x'\cos \omega t+y'\sin\omega t,\qquad t=t',\nonumber \\ 
y & = & r\sin\theta \sin\phi =-x'\sin \omega t+y'\cos\omega t, \qquad z=r\cos\theta =z',
\eea
where $(x^\mu )=(ct,x,y,z)$ and $(x'^\mu )=(ct'=ct,x',y',z'=z)$ are the same coordinates as before, which define the reference frames F (the rotating one) and F$'$ (the inertial one). It results from (\ref{rotating spherical}) that Ryder's tetrad fields (\ref{Ryder 3}) and (\ref{Ryder 10}) verify
\be
\widetilde{u}_0=u_0.
\ee
As noted by Ryder, when the Levi-Civita connection is expressed in the tetrad field (\ref{Ryder 10}), the only non-zero coefficients are $\gamma _{122}=-\gamma _{212}=\gamma _{133}=-\gamma _{313}$ and $\gamma _{233}=-\gamma _{323}$. It thus follows from Eq. (\ref{Phi explicit}) that, in the inertial reference frame F$'$, whose 4-velocity is $v=u_0$ [Eq. (\ref{Minkowski tetrad in rotating Cartesian-0})], the angular velocity tensor field of the spatial triad $({\bf \widetilde{u}}_p)$ taken from the tetrad field $(\widetilde{u}_\alpha )$ is 
\be\label{Xi Ryder2}
\Mat{\widetilde{\Xi}}=\Mat{0}.
\ee
Therefore, here again, we may apply the \hyperref[Comparison]{second paragraph} of this Section. In particular, the two tetrads (\ref{Ryder 3}) and (\ref{Ryder 10}) lead, in the inertial frame F$'$, to non-equivalent Hamiltonian operators. \\

It thus seems that only experiment could tell which of the two natural prescriptions ({\bf i}) and ({\bf ii}) is correct.

\vspace{5mm}
{\bf Acknowledgement.} It is a pleasure to thank Frank Reifler for many interesting discussions on this subject, at an earlier stage of this research.\\


\appendix

\section{Appendix: Geometrical definition of the spatial triad and its angular velocity}\label{Proof1}

\paragraph {Theorem 1.}\label{Theorem1}  {\it Let $(u_\alpha )$ be a global orthonormal tetrad field on the spacetime $(\mathrm{V},\Mat{g})$, so that $v \equiv u_0$ is a global 4-velocity vector field. Assuming that there is at least one chart $\chi $ of $\mathrm{V}$, with domain $\mathrm{U}$, which is adapted to the reference fluid with 4-velocity $v$, let $\mathrm{M}$ be the corresponding space manifold, whose points are the world lines (\ref{world-line}).} (i) {\it At each point $X \in \mathrm{U}$, there is a canonical isomorphism $i_X$ between the three-dimensional vector spaces $\mathrm{TM}_{x(X)}$ and}
\be\label{H_X}
\mathrm{H}_X \equiv \{ u_X \in \mathrm{TV}_X\,;\ \Mat{g}(u_X,v(X))=0 \}.
\ee
(ii) {\it The $\Phi _{pq}$'s $(p,q=1,2,3)$ in Eq. (\ref{Phi ST}) are the components of an (antisymmetric) spatial $(0\ 2)$ tensor field $\Mat{\Phi}$, i.e., a mapping $X \in \mathrm{U} \mapsto \Mat{\Phi}(X) \in \mathrm{TM}^\circ_x \otimes \mathrm{TM}^\circ_x$, with $x=x(X)$. (See Footnote 5.)} (iii) {\it The triad of spatial vector fields $({\bf u}_p)$, with ${\bf u}_p(X)\equiv i_X(u_p(X))$, is orthonormal w.r.t. the spatial metric tensor field $\Mat{h}$ on $\mathrm{M}$.} (iv) {\it We have $\Mat{\Phi}=\Mat{-\Xi}$, where $\Mat{\Xi}$ is the angular velocity tensor (\ref{Xi _pq}) of that orthonormal triad $({\bf u}_p)$  in the space manifold $\mathrm{M}$ equipped with the time-dependent Riemannian metric $\Mat{h}$.}\\

\noindent {\it Proof. } Since the tetrad field $(u_\alpha )$ is orthonormal, each of the three vectors $u_p(X)\ (p=1,2,3)$ at each point $X \in \mathrm{V}$ belongs to the spatial hyperplane (\ref{H_X}). If in fact $X \in \mathrm{U}$, and if $u_X \in \mathrm{H}_X$, let us define $i_X(u_X)$ as the vector ${\bf u}_x \in \mathrm{TM}_{x(X)}$, whose components in the chart $\widetilde{\chi }$ of M [Eq. (\ref{def-chi-tilde})] are $u^j\ (j=1,2,3)$ --- where $u^\mu \ (\mu =0,...,3)$ are the components of $u_X$ in the chart $\chi $. If $\chi '$ is any other chart, with domain $\mathrm{U}$, which is adapted to the same reference fluid, then by definition the transition map $F\equiv \chi '\circ \chi ^{-1}$ satisfies Eq. (\ref{adapted-change}) for ${\bf X} \in \chi (\mathrm{U})$. Hence the spatial coordinates of $u_X$ in the charts $\chi $ and $\chi '$ exchange by
\be
u'^j = \frac{\partial x'^j}{\partial x^\mu }u^\mu =\frac{\partial x'^j}{\partial x^k }u^k,
\ee
thus are the respective components of one and the same vector ${\bf u}_x = i_X(u_X) \in \mathrm{TM}_{x(X)}$ in the charts $\widetilde{\chi} $ and $\widetilde{\chi '}$. Thus, the linear mapping $i_X$ from $\mathrm{H}_X $ to $\mathrm{TM}_{x(X)}$ is well defined. With the $(+---)$ signature, the orthogonal projection operator $\Pi _X: \mathrm{TV}_X \rightarrow \mathrm{H}_X$ \cite{Cattaneo1969,JantzenCariniBini1992,A16} has components 
\be\label{Pi components}
\Pi ^\mu _{\ \, \nu }=\delta ^\mu _{\ \, \nu } -v^\mu v_\nu .
\ee
It follows that, for a vector $u_X \in \mathrm{H}_X$, in any adapted chart, the time component $u^0$ is determined linearly as function of the spatial components $u^j$ by (\cite{A16}, Eq. (4.7)):
\be\label{u^0 for u in H_X}
u^0=-g_{0j}\,u^j/g_{00}.
\ee  
Therefore, if $i_X(u_X)=0$ for some vector $u_X \in \mathrm{H}_X$, then $u_X=0$, so that $i_X$ is indeed an isomorphism. This proves point (i).\\

Using $i_X$, we associate with the spacetime vector fields $u_p\ (p=1,2,3)$, such that $\forall X \in \mathrm{V} \ \ u_p(X) \in \mathrm{H}_X$, the three spatial vector fields 
\be\label{Def bf u_p}
{\bf u}_p:\ X \in \mathrm{U} \mapsto {\bf u}_p(X)\equiv i_X(u_p(X)),
\ee
thus building the {\it spatial triad field} $({\bf u}_p)$ associated with the tetrad field $(u_\alpha )$. Denoting the dual triad of one-forms over the space manifold M as $(\theta^p)$, we define 
\be\label{Phi space}
\Mat{\Phi} \equiv \Phi _{p q} \theta^p \otimes \theta^q.
\ee
Clearly, this is a $(0\ 2)$ tensor field $\Mat{\Phi}$ over the manifold M. This proves point (ii). \\

The spatial metric tensor is first defined as a spacetime tensor $\Mat{h}^\mathrm{(ST)}$, namely as the projection of the spacetime metric tensor \cite{JantzenCariniBini1992,A16}:
\bea\label{Def h^ST}
\forall u_X,v_X \in \mathrm{TV}_X,\quad \Mat{h}_X^\mathrm{(ST)}(u_X,v_X) & \equiv & -\Mat{g}_X(\Pi _X(u_X),\Pi _X(v_X)) \\\nonumber
& = & - \Mat{g}_X(u_X,\Pi _X(v_X)) = -\Mat{g}_X(\Pi _X(u_X),v_X).
\eea
(The two equalities occur because $\Pi _X$ is symmetric w.r.t. $\Mat{g}$ and idempotent. The minus sign is for the $(+ - - -)$ signature of $\Mat{g}$.) Then we define the spatial metric tensor as a truly {\it spatial} tensor field $\Mat{h}$ by transport with the isomorphism $i_X$. Thus, given $X \in \mathrm{U}$, $\Mat{h}_X$ is the mapping
\be\label{Def h}
{\bf u}, {\bf v} \in \mathrm{TM}_{x(X)} \mapsto \Mat{h}_X({\bf u}, {\bf v}) \equiv \Mat{h}_X^\mathrm{(ST)}(i_X^{-1}({\bf u}),i_X^{-1}({\bf v})) = -\Mat{g}_X(i_X^{-1}({\bf u}),i_X^{-1}({\bf v})).
\ee
Since $\Pi_X(u_p(X))=u_p(X) \ (p=1,2,3)$, it follows from (\ref{Def bf u_p}), (\ref{Def h^ST}) and (\ref{Def h}) that (at any $X \in \mathrm{U}$)
\be\label{(u_p) orthonormal}
h_{pq}\equiv \Mat{h}({\bf u}_p,{\bf u}_q) \equiv \Mat{h}^\mathrm{(ST)}(u_p,u_q)=-\Mat{g}(u_p,u_q)=\delta _{p q} \quad (p,q=1,2,3).
\ee
The spatial triad is thus orthonormal for the spatial metric. This proves point (iii).\\

The Fermi-Walker derivative of a spacetime vector field $u$ along a smooth curve $C: \xi \mapsto C(\xi )=X \in \mathrm{V}$, can be defined as the projection of its absolute derivative (\ref{absolute derivative}) on the hyperplane H$_{C(\xi )}$ \cite{JantzenCariniBini1992,A16}: 
\be\label{FW derivative}
\left(\frac{Du}{d\xi }\right )_\mathrm{fw} \equiv \Pi_{C(\xi )} \left(\frac{Du}{d\xi }\right ),
\ee
where the hyperplane and the corresponding projection operator refer to a given 4-velocity field $v$, Eqs. (\ref{H_X}) and (\ref{Pi components}). (We omit the index specifying the curve $C$.) We may naturally infer from this a time-derivative for spatial vector fields $X \mapsto {\bf u}(X) \in \mathrm{TM}_{x(X)}$:
\be\label{delta derivative}
\frac{\delta {\bf u}}{d\xi } \equiv i_{C(\xi )}\left[\left(\frac{Du}{d\xi }\right )_\mathrm{fw}\right ],\quad u(X)\equiv i_X^{-1}({\bf u}(X)).
\ee
It is thus the Fermi-Walker derivative of a true ``spatial vector". The Fermi-Walker derivative of a spacetime vector, Eq. (\ref{FW derivative}), obeys the Leibniz rule with the spatial metric in the spacetime form, $\Mat{h}^\mathrm{(ST)}$, for vector fields such that $u(X) \in \mathrm{H}_X$  (\cite{A16}, Eq. (4.15)). It follows easily from this and from the definition (\ref{Def h}) that the time-derivative (\ref{delta derivative}) obeys also the Leibniz rule:
\be\label{Leibniz for delta}
\frac{d}{d\xi }\,\left(\Mat{h}({\bf u},{\bf u}')\right)=\Mat{h}\left(\frac{\delta {\bf u}}{d\xi },{\bf u}'\right)+\Mat{h}\left({\bf u},\frac{\delta {\bf u}'}{d\xi }\right).
\ee
We may thus transpose Eqs. (\ref{Phi tensor})--(\ref{Phi ST}): We define a spatial tensor by setting
\be\label{Xi tensor}
\frac{\delta {\bf u}_q }{d\xi }  = \Xi ^p _{\ \,q }\, {\bf u}_p;
\ee
since the triad $( {\bf u}_p)$ is orthonormal w.r.t. the metric $\Mat{h}$, we have 
\be\label{Xi _pq}
\Mat{h}\left({\bf u}_p ,\left(\frac{\delta {\bf u}_q }{d\xi }\right) \right)=\Xi ^r  _{\ \,q   }\,\delta _{p r } \equiv \Xi _{p q };
\ee
it follows then from  Eq. (\ref{Leibniz for delta}) with ${\bf u}={\bf u}_p$ and ${\bf u}'={\bf u}_q$ that the spatial tensor $\Mat{\Xi }=\Xi_{p q} \theta^p \otimes \theta^q$ is antisymmetric. Moreover, $\Xi _{pq}=\Xi ^p _{\ \,q }$ from (\ref{Xi _pq}). Thus, the tensor $\Mat{\Xi }$ is the angular velocity tensor of the orthonormal triad $({\bf u}_p)$. From (\ref{FW derivative})--(\ref{delta derivative}), (\ref{Phi tensor}), (\ref{Def bf u_p}), we get (with $X=C(\xi )$):
\bea\label{Phi = Xi}\nonumber
\frac{\delta {\bf u}_q}{d\xi } & \equiv & i_X \left ( \Pi_{X} \left(\frac{Du_q}{d\xi }\right )\right )=i_X \left (\Pi_{X} \left ( \Phi ^\alpha _{\ \, q} u_\alpha \right )\right )=i_X\left( \Phi ^p _{\ \, q} u_p \right )\\
& = & \Phi ^p _{\ \, q} {\bf u}_p,
\eea
so that in fact $\Xi ^p _{\ \,q } = \Phi ^p _{\ \,q }$. However, from (\ref{Phi ST}) we have 
\be
\Phi_{pq}\equiv \,\eta _{p \epsilon } \Phi ^\epsilon _{\ \,q }= -\Phi ^p _{\ \,q },
\ee
hence $\Phi_{pq}=-\Xi_{pq}$, or $\Mat{\Phi }=\Mat{-\Xi }$. The results of this paragraph apply to any curve $C$ in spacetime, with any parameter $\xi $. To complete the proof of \hyperref[Theorem1]{Theorem 1}, we need that the tensor $\Mat{\Phi }=\Mat{-\Xi }$ be a field defined on U. At any $X \in \mathrm{U}$, we take the world line $x(X)$, parameterized by the time coordinate $t$ as in Eq. (\ref{w=v ds/dt}), as the curve $C$. This proves point (iv). 

\section{Appendix: Rotating Cartesian coordinates}\label{RotFlat}

The cotetrad and tetrad used by Ryder \cite{Ryder2008} in his equations (1) and (3), respectively, are thus given in coordinates $t,x,y,z$ which are related to Cartesian coordinates $t'=t,x',y',z'=z$ in an inertial frame in a Minkowski spacetime by a rotation with angular velocity $\Mat{+}\omega $. To check this, consider the line element \cite{L&L-89}
\be\label{L&L(89,2)}
ds^2=\left[1-\left(\frac{\omega \rho }{c}\right)^2\right]c^2 dt^2 - 2\omega \rho ^2 \,d\varphi\, dt-(d\rho ^2+\rho ^2 d\varphi ^2+dz^2),
\ee
deduced from the Minkowski line element in cylindrical coordinates:
\be\nonumber
ds^2=c^2 dt'^2 -(d\rho '^2+\rho '^2 d\varphi'^2+dz'^2),
\ee
by the coordinate change $t=t',\,\rho =\rho' ,\,\varphi =\varphi' -\omega t,\,z=z'$.
This corresponds to a rotation with angular velocity $\Mat{+}\omega $ w.r.t. the inertial frame with cylindrical coordinates $t',\,\rho ',\,\varphi ',\,z'$, because $\varphi '=\varphi +\omega t$ is indeed the polar angle in the inertial coordinates, corresponding to the polar angle $\varphi $ in the rotating coordinates. Introducing ``rotating Cartesian coordinates":
\be\label{rotating Cartesian}
t,\quad x=\rho  \cos\varphi ,\quad y=\rho  \sin\varphi ,\quad z,
\ee
and noting that
\be
x\, dy-y\,dx=(\rho ^2 \cos^2\varphi -(-\rho ^2 \sin^2\varphi))d\varphi =\rho ^2 d\varphi,
\ee
the line element (\ref{L&L(89,2)}) rewrites as Eq. (\ref{Minkowski in rotating Cartesian}):
\be\label{Minkowski in rotating Cartesian-2}
ds^2=\left[1-\left(\frac{\omega }{c}\right)^2\,(x^2+y^2)\right]c^2 dt^2 + 2\omega (y\,dx-x\, dy)\, dt-(dx^2+dy^2 +dz^2),
\ee
which is the first line element used by Ryder \cite{Ryder2008} (with a misprint). One checks trivially that the cotetrad (1) of Ryder is orthonormal for the metric (\ref{Minkowski in rotating Cartesian-2}) above, i.e., we have with the metric (\ref{Minkowski in rotating Cartesian-2}):
\be
ds^2\equiv g_{\mu \nu }dx^\mu\, dx^\nu = \eta _{\alpha \beta }\, \theta ^\alpha \,\theta ^\beta .
\ee
A point with constant spatial coordinates $x,y,z$ given by (\ref{rotating Cartesian}), with $\varphi=\varphi' -\omega t$, is indeed rotating at $+\omega $ w.r.t. the inertial frame defined by the Cartesian coordinates $t'=t,x',y',z'=z$. This is because we get from (\ref{rotating Cartesian}) the relevant formula (\ref{rotating Cartesian-1}):
\bea\nonumber
x & = &\rho  \cos(\varphi'-\omega t)= \rho (\cos\varphi '\cos\omega t+\sin\varphi '\sin\omega t)\equiv x'\cos\omega t+y'\sin\omega t,\\
\nonumber
y & = & \rho  \sin(\varphi'-\omega t)= \rho (\sin\varphi '\cos\omega t-\cos\varphi '\sin\omega t)\equiv -x'\sin\omega t+y'\cos\omega t. 
\eea

\section{Appendix: Analysis of the proposals of Gorbatenko \& Neznamov}\label{GorbatNeznam}

For a general, time-dependent metric, Gorbatenko \& Neznamov \cite{GorbatenkoNeznamov2011} want to define a unique Hermitian Hamiltonian operator $\mathrm{H}_\eta$ by going from the starting tetrad field with components $H^\mu _{\underline{\alpha }}$ (used to define the starting field of Dirac matrices noted $\gamma ^\alpha  $) to what they call ``the Schwinger tetrad", i.e., one whose components $\widetilde{H}^\mu _{\underline{\alpha }}$ (in the chart which is considered) satisfy $\widetilde{H}^0 _{\underline{k }}=0 \quad (\underline{k }=1,2,3)$, their Eq. (39). With the Schwinger tetrad, the field of Dirac matrices is $\widetilde{\gamma} ^\alpha  $. Gorbatenko \& Neznamov denote by $L$ a local similarity transformation which transforms $\gamma ^\alpha  $ into $\widetilde{\gamma} ^\alpha  $, their Eq.  (52). They compute from $L$ another local similarity transformation $\eta $, Eqs. (66) and (72). The Hamiltonian operator which rewrites in Schr\"odinger form the covariant Dirac equation after the local similarity $\eta $, is named $\mathrm{H}_\eta$, Eq. (71). When the starting tetrad $H^\mu _{\underline{\alpha }}$ is taken to be ``the Schwinger tetrad" $\widetilde{H}^\mu _{\underline{\alpha }}$, we have of course $L={\bf 1}_4$. Then, this process defines a local similarity $\widetilde{\eta }$ proportional to the identity matrix, Eq. (67), and a Hamiltonian, say $\widetilde{\mathrm{H}}_{\widetilde{\eta }}$. Gorbatenko \& Neznamov \cite{GorbatenkoNeznamov2011} correctly note that $\mathrm{H}_\eta=\widetilde{\mathrm{H}}_{\widetilde{\eta }}$, Eq. (75). In fact, it is easy to check that $\widetilde{\mathrm{H}}_{\widetilde{\eta }}$ is none other than the energy operator [Eq. (\ref{E:=H^s}) in the present paper] got with the Schwinger tetrad $\widetilde{H}^\mu _{\underline{\alpha }}$, i.e., $\mathrm{H}_\eta=\widetilde{\mathrm{H}}_{\widetilde{\eta }}$ is the energy operator with the Dirac matrices $\widetilde{\gamma} ^\alpha  $.  (This is confirmed by the recent paper \cite{GorbatenkoNeznamov2011b}: Eq. (1) there.) Gorbatenko \& Neznamov \cite{GorbatenkoNeznamov2011} conclude that ``For any system of tetrad vectors, after the specified operations we will have one and the same Hamiltonian $\mathrm{H}_\eta$ (75), what proves its uniqueness." However, both $\mathrm{H}_\eta$ and $\widetilde{\mathrm{H}}_{\widetilde{\eta }}$ are defined with reference to one given Schwinger tetrad. But the choice of a Schwinger tetrad is far from unique. Gorbatenko \& Neznamov do not prove that, if in the same chart one takes a second Schwinger tetrad, say $\breve{H}^\mu _{\underline{\alpha }}$, then the Hamiltonian provided by their construction for that second Schwinger tetrad, say $\breve{\mathrm{H}}_{\breve{\eta }}$, is equivalent to the Hamiltonian $\widetilde{\mathrm{H}}_{\widetilde{\eta }}$ got with the first Schwinger tetrad. \\

Coming back to our notation for convenience, let $(u_\alpha )$ be a tetrad field which is a Schwinger tetrad in a given chart $\chi :X\mapsto (x^\mu )$. That is:
\be\label{a^0_p=0}
a^0_{\ \,p}\equiv (u_p)^0=0 \quad (p=1,2,3).
\ee
[Note from (\ref{purely-spatial-change}) that this remains true for any chart $\chi '$ in the reference frame F defined by $\chi$.] If one changes by a local Lorentz transformation to a new tetrad: $\widetilde{u}_\beta \equiv L^\alpha _{\ \, \beta }\,u_\alpha $, then this will be also a Schwinger tetrad in the same reference frame F iff
\be
\widetilde{a}^0_{\ \,p}\equiv L^\alpha _{\ \, p }\,a^0_{\ \,\alpha }=0 \quad (p=1,2,3),
\ee
hence, in view of (\ref{a^0_p=0}), iff
\be\label{L^0_p=0}
L^0 _{\ \, p }=0 \quad (p=1,2,3).
\ee
This does not require that the local Lorentz transformation $L$ be independent of the time coordinate $t$, nor hence that the associated local similarity $S=\pm {\sf S}(L)$ verify $\partial _t S=0$ [condition (\ref{partial_0 S=0})], so that the two Dirac Hamiltonians deduced from the tetrad fields $(u_\alpha )$ and $(\widetilde{u}_\alpha )$ can be non-equivalent. 
\\

For instance, in the {\it inertial frame} F$'$ in a {\it Minkowski spacetime,} considered in Subsect. \ref{Omega} and in Sect. \ref{Discussion}, take Ryder's first tetrad $(u_\alpha )$, Eq. (\ref{Ryder 3}), and take the ``Minkowski tetrad", that is the natural basis $(e'_\mu )$ associated with the Cartesian coordinates $(x'^\mu )=(ct',x',y',z')$ which define the reference frame F$'$. Using the coordinates $(x^\nu )=(ct=ct',x,y,z=z')$, Eq. (\ref{rotating Cartesian-1}), which define the uniformly rotating reference frame F, we may write:
\be
e'_\mu \equiv \frac{\partial }{\partial x'^\mu }=\frac{\partial x^\nu }{\partial x'^\mu }\frac{\partial }{\partial x^\nu },
\ee
hence from (\ref{rotating Cartesian-1}):
\bea\label{Minkowski tetrad in rotating Cartesian-0}
e'_0 & = & \frac{\partial }{\partial x^0 }+\frac{\omega y}{c}\frac{\partial }{\partial x}-\frac{\omega x}{c}\frac{\partial }{\partial y} = u_0,
\\
\label{Minkowski tetrad in rotating Cartesian-123}
e'_1 & = & \cos \omega t \,u_1-\sin \omega t\,u_2,\quad e'_2=\sin \omega t \,u_1+\cos \omega t\,u_2, \quad e'_3=u_3.
\eea
The Minkowski tetrad $(e'_\alpha )$ is a Schwinger tetrad in the inertial frame F$'$, hence by inverting (\ref{Minkowski tetrad in rotating Cartesian-123}) we see that Ryder's first tetrad $(u_\alpha )$ is also a Schwinger tetrad in that frame F$'$. (Note that Eqs. (\ref{Minkowski tetrad in rotating Cartesian-0})--(\ref{Minkowski tetrad in rotating Cartesian-123}) allow us to define $(u_\alpha )$ in the {\it global} frame F$'$.) Therefore, with either of these two tetrad fields, the hermiticity condition of the Dirac Hamiltonian in the frame F$'$ is Leclerc's condition \cite{Leclerc2006}:
\be\label{Leclerc-2}
\partial _0 \left(\sqrt{- \,g^{00}\,\mathrm{det}(g_{\mu \nu })}\right)=0
\ee
(provided that one takes for the ``flat" Dirac matrices $\gamma ^{ \natural \alpha}$ in Eq. (\ref{flat-deformed}), ones for which $\gamma ^{ \natural 0}$ is a hermitizing matrix \cite{A42}). Since the Minkowski metric is stationary in the inertial frame F$'$, the condition (\ref{Leclerc-2}) is satisfied in F$'$. Thus, each of the Dirac Hamiltonians H and H$'$ associated respectively, in the frame F$'$, with the Schwinger tetrads $(u_\alpha )$ and $(e'_\alpha )$, coincides with the corresponding energy operator: $\mathrm{H=E}$ and $\mathrm{H'=E'}$. I.e., H (respectively H$'$) is the Hermitian Hamiltonian associated with the Schwinger tetrad $(u_\alpha )$ [respectively the Schwinger tetrad $(e'_\alpha )$] by the procedure suggested by Gorbatenko \& Neznamov \cite{GorbatenkoNeznamov2011}. However, from (\ref{Minkowski tetrad in rotating Cartesian-0})--(\ref{Minkowski tetrad in rotating Cartesian-123}), we have $e'_\beta = L^\alpha _{\ \, \beta }\,u_\alpha $, with the {\it time-dependent} local Lorentz transformation
\be\label{L for Ryder1 and Minkow}
L = \begin{pmatrix} 
1 & 0  & 0 & 0\\
0  & \cos \omega t & \sin \omega t & 0\\
0 & -\sin \omega t & \cos \omega t & 0\\
0 & 0  & 0 & 1
\end{pmatrix}.
\ee
Hence, the associated local similarity $S=\pm {\sf S}(L)$ does not verify $\partial _t S=0$, so that the Hamiltonians H and H$'$ are not equivalent. However, each of the two non-equivalent Hamiltonians H and H$'$ is got from a tetrad which is a Schwinger tetrad in the given coordinates $(x'^\nu )=(ct',x',y',z')$, by following the procedure suggested by Gorbatenko \& Neznamov \cite{GorbatenkoNeznamov2011}.\\

In their most recent paper \cite{GorbatenkoNeznamov2011b}, Gorbatenko \& Neznamov included in the definition of a ``Schwinger tetrad" two additional conditions (with our notation and signature): {\bf a}) that the time-like vector $u_0$ of the tetrad be the vector with components
\be\label{u_0 for prescr. a}
a^0 _{\ \,0}=\sqrt{g^{00}},\quad a^k _{\ \,0}=\frac{g^{0k}}{\sqrt{g^{00}}},
\ee
Eq. (15) in Ref. \cite{GorbatenkoNeznamov2011b}. And {\bf b}) that the spatial components of the spatial triad $(u_p)$ verify [their Eq. (19)]
\be\label{spatial triad G&N}
a^j _{\ \,p}\,a^k _{\ \,p}=f^{jk}, \quad \mathrm{or}\quad a\, a^T = f \quad [a\equiv (a^j _{\ \,p}), \quad f\equiv (f^{jk})],
\ee
where $f$ is the inverse of the $3\times 3$ matrix $-g\equiv -(g_{jk})$, their Eq. (18). Equation (\ref{spatial triad G&N}) is equivalent to $a^T\,(-g)\,a={\bf 1}_3$. It thus means that the spatial vector fields ${\bf u}_p(X)$, with components $a^j _{\ \,p}(X)$, make an orthonormal triad for the ``induced metric" $\Mat{h}'\equiv -g_{jk} dx^j \otimes dx^k$. Actually, both (\ref{u_0 for prescr. a}) and (\ref{spatial triad G&N}) follow from the definition (\ref{a^0_p=0}) of a Schwinger tetrad and the orthonormality condition (\ref{orthonormal tetrad}) of the tetrad. (This was indeed noted by the authors, respectively before Eq. (79) of Ref. \cite{GorbatenkoNeznamov2011}, and as Eqs. (50)--(51) of Ref. \cite{GorbatenkoNeznamov2011b}v1, but it escaped our attention for v1 of the present paper.) However, just as we did with the other spatial metric $\Mat{h}$ defined by Eq. (\ref{Def h}), we may define different such triad fields, say $({\bf u}_p)$ and $({\bf \widetilde{u}}_p)$, {\it having an arbitrary rotation rate w.r.t. one another.} In Sect. 4 of Ref. \cite{GorbatenkoNeznamov2011b} (in v1 but not in v2 and v3), yet another condition was added: {\bf c}) that the spatial vector fields ${\bf u}_p(X)$ be eigenvectors of the metric tensor $\Mat{h}'$ [$g_{mn}$ in their notation]. The motivation for imposing the Schwinger-tetrad condition (\ref{a^0_p=0}), or for imposing condition {\bf c}), looks computational rather than physical. Anyway, these two conditions together do not forbid to define triad fields $({\bf u}_p)$ and $({\bf \widetilde{u}}_p)$ which exchange by a time-dependent rotation, if some eigenvalues of $\Mat{h}'$ are equal. Both Ryder's first tetrad $(u_\alpha )$ and the ``Minkowski tetrad" $(e'_\alpha )$ satisfy the three additional conditions {\bf a}), {\bf b}) and {\bf c}) in the Cartesian coordinates $(x'^\mu )$: condition {\bf c}) is trivial here, since $h'_{jk}=\delta _{jk}$; and as it has been noted, conditions {\bf a}) and {\bf b}) follow generally from (\ref{a^0_p=0}) and (\ref{orthonormal tetrad}). Hence, the same counterexample detailed around (\ref{L for Ryder1 and Minkow}) applies also with the additional conditions {\bf a})-{\bf b})-{\bf c}). 


\end{document}